# Profiling underprivileged residents with mid-term public transit smartcard data of Beijing

Ying Long, Beijing Institute of City Planning, China, longying1980@gmail.com
Xingjian Liu, University of North Carolina at Charlotte, USA
Jiangping Zhou, Iowa State University, USA
Yizhen Gu, University of California, Berkeley, USA

**Abstract:** Mobility of economically underprivileged residents in China has seldom been well profiled due to privacy issue and the characteristics of Chinese over poverty. In this paper, we identify and characterize underprivileged residents in Beijing using ubiquitous public transport smartcard transactions in 2008 and 2010, respectively. We regard these frequent bus/metro riders (FRs) in China, especially in Beijing, as economically underprivileged residents. Our argument is tested against (1) the household travel survey in 2010, (2) a small-scale survey in 2012, as well as (3) our interviews with local residents in Beijing. Cardholders' job and residence locations are identified using Smart Card Data (SCD) in 2008 and 2010. Our analysis is restricted to cardholders that use the same cards in both years. We then classify all identified FRs into 20 groups by residence changes (change, no change), workplace changes (change, no change, finding a job, losing a job, and all-time employed) during 2008-2010 and housing place in 2010 (within the fourth ring road or not). The underprivileged degree of each FR is then evaluated using the 2014 SCD. To the best of our knowledge, this is one of the first studies for understanding long- or mid-term urban dynamics using immediate "big data", and also for profiling underprivileged residents in Beijing in a fine-scale.

## 1 Introduction

China has witnessed rapid economic development in recent decades since the country's reforming and opening-up in the 1980s. Despite the overall economic development, urban poverty has increasingly plagued Chinese cities, among which the post-Olympic Beijing would not be the exception (Song et al., 2009; Wu, 2010). A better understanding of identities and social-spatial mobilities of underprivileged residents with relatively low income is crucial for designing public policies and social programs, such as e.g. affordable housing and public transit system. In this paper, we aim to provide a method for identifying socially and spatially underprivileged residents in the Chinese capital Beijing and detailing their social and spatial mobility based on increasingly available smartcard data from public transit system.

It is challenging to identify these underprivileged residents by conventional methods such as questionnaires (online or offline), face-to-face interview and telephone interview. One major obstacle remains that personal fortune – wealth or poverty – is still a taboo subject in the Chinese culture. Consequently, people's revealed attitudes and/or conditions about poverty is oftentimes not a robust measurement. For instance, the Household Travel Survey of Beijing suggests the overall annual household income in 2010 is around 30,000 CNY (less than 5,000 USD), which is a far cry from numbers coming from other sources, such as other academic studies, statistical yearbooks, and local tales. In addition, poverty conditions assessed via surveys and interviews are often over-stated, as these surveys and interviews are often used to gather information for welfare programs: the temptation to over-stress your need for fiscal help is a strong one. Still, conventional methods such as survey suffer from issues such as limited sample size and coarse spatial resolution. For these reasons, relevant official reports are prone to unrobust measurement and deliberate manipulation. To summarize, the identification of underprivileged residents and their social and spatial mobility is of significant social value, however the operationalization by conventional methods is difficult.

With the development of information and communication technologies (ICT), voluminous, longitudinal, and ubiquitous digital measurements – often dubbed as "big data" emerge as an alternative solution to figure out our research questions through sensing and diagnosing cities (Batty, 2012). Human mobility and activity could be captured by big data, providing detailed spatial and social information in a fine scale for academic research (Goodchild, 2007). Among all types of big data such as GPS trajectories, social networks and phone call records, smartcard data (SCD) have been widely used in public transit systems in China.

Such public transit SCD are a promising candidate for charting the social and economic geography of underprivileged residents for the following reasons. **First**, bus/metro riders in Beijing are generally the low-income group. Most residents who could afford private cars would commute by driving, instead of by taking oftentimes over-crowed public transit. Economic-wise, car ownership is both a positional good as well as a life necessity in the increasingly car-based living in Beijing. Even for those who cannot afford automobiles but still with relative better means, taxis and chartered buses are the more preferred modes of transportation. After all, those who spend hours every day on the city's overcrowd transit system (Figure 1) oftentimes have no other options. **Second**, over 90% of all public transit passengers (mode share in 2010 is 40%) possess identifiable smartcards, as regular card users receive a 60% discount on the fares. Given the large size of public transit riders and the high percentage of smartcard ownership, SCD would indeed cover a sheer population: There are over 10 million cardholders in 2010. **Third**, a person who rides bus/metro for a relatively long period, say for example two years, would have a higher probability to be an underprivileged resident in Beijing: long-time public transit riders often have limited social mobility, as in the case of Beijing This would be our highlight in this paper. To sum up, public transit smart cardholders would overlap with the most underprivileged residents in Beijing.

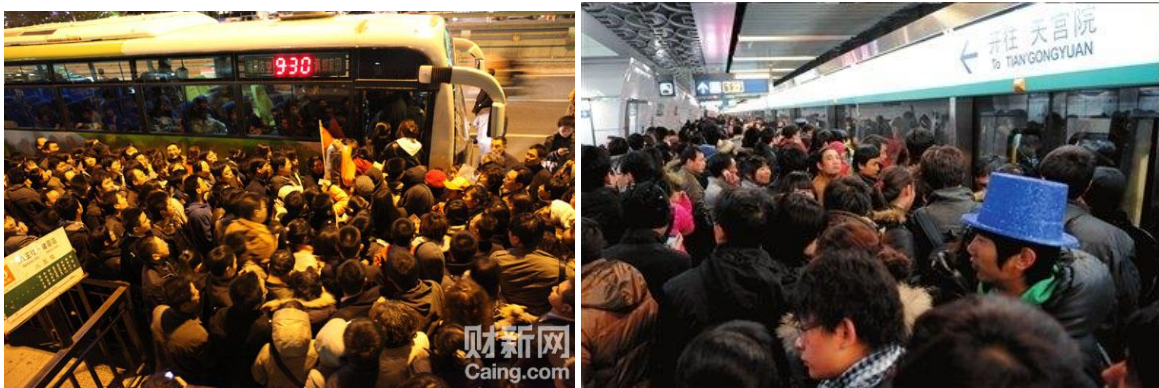

Figure 1 Crowd for riding bus (left, Line 930) and metro (right, Line 4) in afternoon peak of Beijing
Source: www.caing.com

With the advantages of SCD in mind, we should be cautious about some potential bias incurred from treating all frequent cardholders as underprivileged residents. For instance, several colleagues of the first author would commute everyday by bus, in order to live in the downtown and stay close to their jobs. However, the colleagues of the first author would by no means be categorized as second-class residents in Beijing. Therefore, only a portion of cardholders meets the definition of socially and spatially underprivileged residents. Additional rules are needed to separate these residents from the general pool of cardholders. In addition, long-term observations would help to circumvent issues arising from short term fluctuations in ridership. This is consistent with Batty's (2013) claim that it is possible to identify long-term dynamics using intermediate big data.

In this paper, we identify cardholders' places of residence and work using SCD in 2008 and 2010, respectively. Their residential and job location changes could be derived via comparing the results of both years, since every smartcard has a unique ID which remains the same during the study period. The cardholders with identified housing place in both years are taken as underprivileged residents in Beijing, which were then profiled into various groups based on travel frequency, commute time, as well as jobs-housing location and their change. The human mobility of final identified cardholders associated with lower social and spatial mobility is then depicted in fine spatial scale based on which corresponding policy implications are drawn. This paper is structured as follows. Section 2 reviews existing research in both urban poverty and SCD application. Study area, data, and underprivileged residents in Beijing are explained in Section 3. The methods and their results are described in Section 4 and 5, respectively. We draw conclusions and propose potential applications of the study in Section 6.

## 2 Related research

### *2.1 Urban poverty of Chinese cities*

Underprivileged residents have attracted extensive attention from scholars, in the form of research on urban poverty, urban migration, urban villages etc. Almost all existing studies were conducted using small-scale surveys or focus groups (e.g. several-hundred interviewers in several typical neighborhoods) (Chen et al, 2006; Fan et al, 2011; Wu et al, 2010; Zheng et al, 2009). Most of these studies focus on interviewers' living condition, traffic expenditure, public service requirements as well as commuting conditions, rather than residential and work place variation and detailed mobility status, which are not easy to measure by conventional surveys. However, to the best of our knowledge, there is no study that identifies large-scale underprivileged residents, analyzes their jobs-housing variation during a middle term, and visualizes their mobility in a very detailed manner.

## *2.2 Social-economic level identification using trajectories*

There are few studies on automatically identifying social-economic levels of travelers (SELs) using trajectories like mobile traces, SCD, and floating car trajectories. Most studies focus on correlating SELs with large-scale-trajectory-extracted human mobility. For instance, Amini et al (2014) analyzed the impact of developing level on human mobility using 150-day mobile call records. Frias-Martinez et al (2012) studied the relationship between mobility variables and SELs using cell phone traces over six months, and found that populations with higher social-economic levels are strongly linked to larger mobility ranges than populations from lower socio-economic status (Aggregated correlation, not identification from individual mobility). One exception is the research of Kang et al (2010), which used mobile call records of xx months, found those users with fewer but frequently-visited anchor points are associated with stable jobs, while those with more but less frequently-visited anchor points with unstable jobs. This is proven by common sense of Chinese residents as discussed in the introduction section. Therefore, identifying social-economic levels according to the jobs-housing place variation provides an alternative solution.

There are considerable publications on inferring housing and job places from trajectories like mobile phone call data records and location-based social networks (LBSN). For housing place identification, Lu et al (2013) regarded the location of the last mobile signal of the day as the housing place of a mobile user. The most frequently visited point-of-interest (POI) (Scellato et al, 2011)) or grid (Cheng et al, 2011; Cho et al, 2011) was regarded as an LBSN user's housing place. It is not easy to infer housing places form LBSN with a high spatial resolution. Comparing housing place identification, there are fewer studies on identifying job places using trajectories, with the exceptions of Cho et al (2011) using LBSN and Isaacman et al (2011) using cellular network data. It should be mentioned that taxi trajectories are not suitable for identifying a passenger's housing and job places considering the passenger-sharing nature of taxis. However, less attention has been paid to using SCD to identify housing and job places as well as their variation across time in a metropolitan city, which is the main focus of this paper.

## *2.3 Smartcard data mining*

The smartcard recording all cardholders' bus trip information is an alternative form of location-acquisition technology. Smartcard-automated fare collection systems are increasingly applied in public transit systems. Simultaneously with collecting fares, such systems can produce intensive travel patterns of cardholders, data that are useful for analyzing urban dynamics. Since 1990 the use of smartcards has become significant owing to the development of the Internet and the increased complexity of mobile communication technologies (Blythe, 2004). Intelligent Transportation Systems (ITS) that incorporate smartcard-automated fare systems either existed or had been being established in over 100 Chinese cities as of 2007 (Zhou, 2007). The data generated by smartcard systems track the detailed onboard transactions of each cardholder. We argue that smartcard technology can provide valuable information because it is a continuous data collection technique that provides a complete and real-time bus travel diary for all bus travelers.

Previous studies advocate using SCD to make decisions on the planning and design of public transit systems (see Pelletier et al (2011) for a review) as well as to analyze urban structure. In South Korea, Joh and Hwang (2010) analyzed cardholder trip trajectories using bus smartcard data from ten million trips of four million individuals, and correlated this data with land use characteristics in the Seoul Metropolitan Area. Jang (2010) estimated travel time and transfer information using data of more than 100 million trips in Seoul from the same system. Roth et al (2011) used a real-time "Oyster" card database of individual traveler movements in the London subway to reveal the polycentric urban structure of London. Gong et al (2012) explored spatiotemporal characteristics of intra-city trips using metro SCD of 5 million trips in Shenzhen, China. Sun et al (2012) used subway SCD in Singapore to extract passengers' spatio-temporal density and train's trajectory. Sun et al (2013) analyzed encounter patterns of cardholders using around 20 million bus SCD from 2.8 million anonymous users over one week in Singapore. Long and Thill (2013) mapped commuting pattern of Beijing using one-week SCD in 2008 via identifying cardholders' housing and job places. Zhou and Long (2014) used the over 200 thousand identified commuting trips in Long and Thill (2013) to analyze bus commuters' jobs-housing balance in Beijing. In sum, considering urban dynamics process in terms of different response time and duration proposed by Wegener et al (1986) and strengthened by Simmonds et al (2011), existing research focuses on using a short period SCD for instant city mapping or revealing instant urban dynamics, but few of them is related with identifying medium- or long-term urban dynamics (e.g. land use change, residential construction or economic change) using a long-period SCD, with an exception of Sun et al (2014).

> *"…… the power of big data is that if collected for long enough then the longer term will emerge from the short term. At the moment these data are about what happens in the short term, but over ten years or longer we will have a unique focus on the longer term – in fact we will have a snapshot of urban dynamics which is unprecedented."*
>
> Batty, 2013

It is possible to identify medium- or long-term (medium or low-speed) dynamics using intermediate big data. In this paper, housing or job place change for a person belongs to medium-term urban dynamics, which is not easy to analyze using short-term big data. To our knowledge, this manuscript would be the first exploration in this field, in the context that all studies on big data focus on instant dynamics. We argue that, long term dynamics is possible to be identified using big data expanding several years.

# 3 Study area, data and local background

## *3.1 Study area*

As the capital of China, the Beijing Metropolitan Area (BMA) with an area of 16,410 km$^2$ has over 20 million residents in 2010 and is becoming one of the most populous cities in the world. The BMA lies in northern China, to the east of the Shanxi altiplano and south of the Inner Mongolian altiplano. The southeastern part of the BMA is a flatland, extending east for 150 km to the coast of the Bohai Sea. Mountains cover an area of 10,072 km2, 61% of the whole study area (Figure 1). See Yang et al. (2011) for more background information on Beijing.

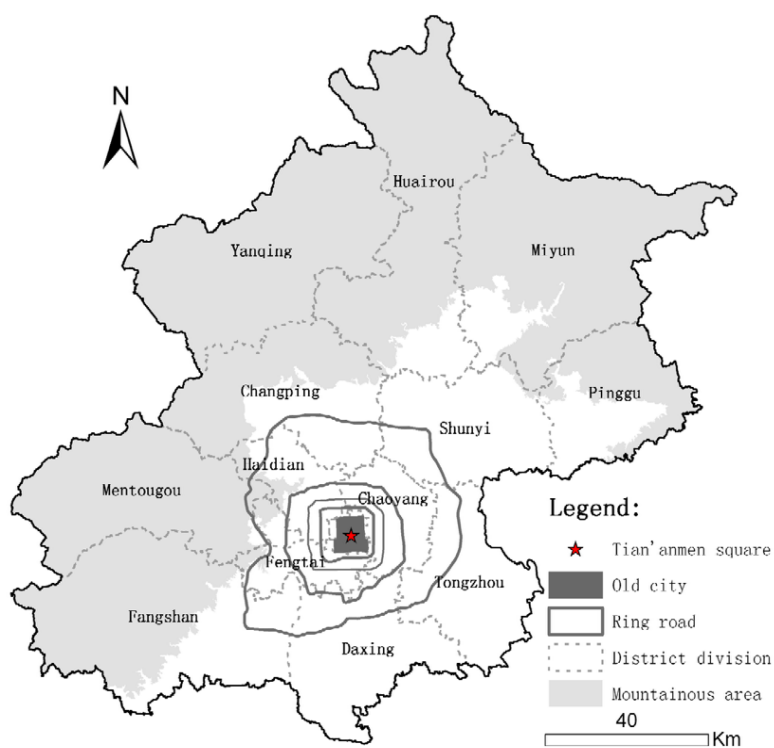

Figure 2 The Beijing Metropolitan Area map

As of 2010, there were 184 kilometers of commuter subway lines (excluding the airport express rail) in Beijing. Beijing Metro manages and maintains the subway system, which was built and financed by the BMG. Beijing Public Transportation Company, a state-owned company provides public bus services in the Beijing metropolitan area. In 2011 alone, these buses produced 1.7 billion of vehicle kilometers traveled and transported total passenger of 4.9 billion[1]. Bus trips accounts for a significant share of all the trips by public transit (subway, bus and company shuttle). Thanks to the continuous expansion of the subway lines in Beijing, the share of subway trips have gradually increased in the city. Table 1 shows the mode share of the local residents of Beijing in 2008 and in 2010.

---

[1] Information based on: http://www.bjbus.com/home/view_content.php?uSec=00000002&uSub=00000012, accessed July 01, 2012.

Table 1: Mode Share of Beijing Residents

| Mode | 2008 Share (%) | 2010 Share (%) |
|---|---|---|
| Bus | 28.8 | 28.9 |
| Metro | 8.0 | 10.0 |
| Taxi | 7.4 | 7.1 |
| Car | 33.6 | 34.0 |
| Bike and walking | 20.3 | 18.1 |
| Company shuttle | 1.9 | 1.9 |
| Total | 100 | 100 |

Sources: Beijing Transportation Research Center, 2011

## *3.2 Data*

### 3.2.1 Smartcard data in 2008 and 2010

Since 2005, over 90% of bus riders in Beijing have swiped an anonymous smartcard (images see Figure 3) when boarding and alighting (for suburb routes) or when boarding (for inner-city routes) to pay for their fare. The high rate of smartcard usage among bus riders is largely because of the subsidy the government gives to riders paying the bus fare by a smartcard. Those riders enjoy 60% discounts on any routes in the local bus system (80% for students). Smartcard also enables owners to pay for other services such as taxi, electricity and sewage that are offered by the local government or companies it subsidizes.

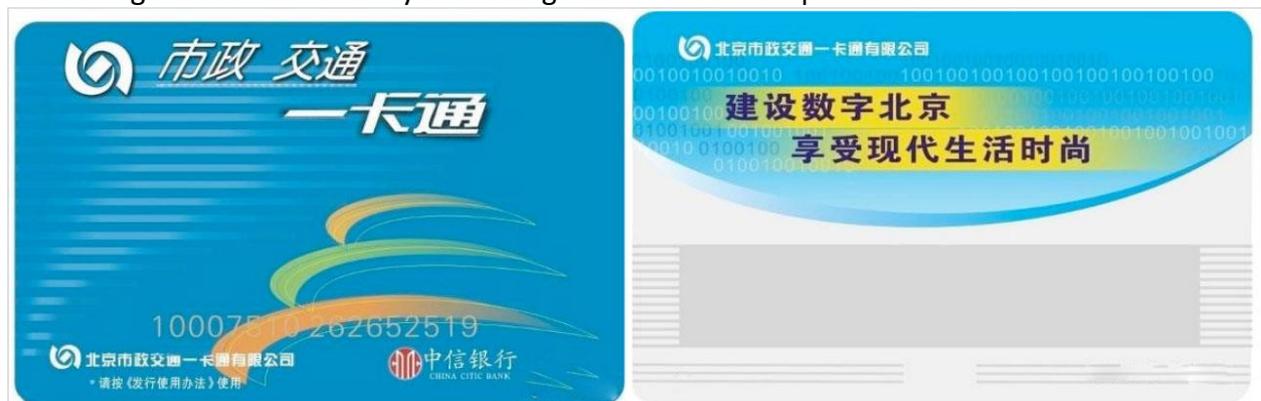

Figure 3: A public transit smartcard in Beijing (Left: front cover, Right: back cover)

When cardholders use their smartcard for paying bus services, the card reader installed on the bus automatically generates the following information, Bus trip origin and/or destination stops, boarding and/or alighting time, as well as the unique card number and the card type (student card at a discount vs. regular card). Two bus fare types exist. The first is fixed-fare, which is associated with short routes, and the other is distance-fare, which is associated with long routes. For the first type, 0.4 CNY is charged for each single bus trip, and the corresponding SCD contains only the departure time and stop ID and no arrival time or stop ID. Cardholders' spatiotemporal information is incomplete for this kind of route. For the latter type, the fare depends on the route ID and trip distance, and the SCD contains full information. Both types of bus SCD records are used in this paper for identifying cardholders' housing and job places as well as residents that are socially and spatially underprivileged.

Our strategy for tackling incomplete information issue of fixed-fare records have been elaborated in Long and Thill (2013), and the strategy proposed by Ma et al (2012) could be used in future for solving the incomplete information of Beijing bus SCD. The records of subways have complete spatiotemporal information, the same as those of flexible-fare bus records.

We collected one-week SCD in 2008 and 2010 from Beijing Municipal Administration & Communications Card Co. (BMAC). They are the key data of this paper. Details are in Table 2. Subway records are missing for 2008 SCD and included in 2010 SCD. SCD have been geo-coded using bus route and stops/stations in Figure 4, which is a time consuming process.

Table 2: The summary of SCD in 2008 and 2010

| Year | 2008 | 2010 |
|---|---|---|
| Date | 7-13 April | 5-11 April |
| # cardholder (m) | 8.5 | 10.9 |
| # SCD (m) | 78.0 | 97.9 |
| 1 # Bus (m) | 78.0 | 82.7 |
| 1.1 # Flexible fare (m) | 27.0 (34.7%) | 23.4 (28.3%) |
| 1.2 # Fixed fare (m) | 51.0 (65.3%) | 59.3 (71.7%) |
| 2 # Subway (m) | 0 | 15.2 |

**3.2.2 Bus routes/stops, subway lines/stations, and traffic analysis zones (TAZs) of Beijing**

GIS layers of bus routes and stops and subway lines and stops are essential for geocoding and mapping SCD. Since SCD in two years are included in this paper, and bus routes and stops change a lot during 2008-2010, we used different spatial layers to map SCD in 2008 and 2010, respectively. In 2008, there were 1,287 bus routes (Figure 1a) and 8,691 bus stops (see Figure 1b) in the BMA. In 2010, there were 1928 bus routes (Figure 1c) and 21,372 bus stops (Figure 1d) in the BMA. Note that a bus route in this paper has direction, e.g., the bus No 113 has two routes, one from Dabeiyao to Qijiahuozi and the other from Qijiahuozi to Dabeiyao. They are counted separately in this paper. In addition, a pair of bus platforms on opposite sides of a street is considered as one bus stop. For instance, there are two bus platforms at Tian'anmen Square, one on the south side of Chang'an Avenue and the other on the north side. In the bus stop GIS layer, the two platforms are merged into one bus stop feature. The average distance between each bus stop and its nearest neighbor in 2010 is 231 m. Since SCD in 2010 include subway information, subway lines and stations are used to map SCD2010. In 2010, there are 9 subway lines including the airport express line and 147 subway stations associated with the 2010 SCD (see Figure 1e). We use Beijing TAZ data to aggregate the analytical results for better visualization. Totally 1,911 TAZs are defined (see Figure 1f) according to the administrative boundaries, main roads, and the planning layout in the BMA.

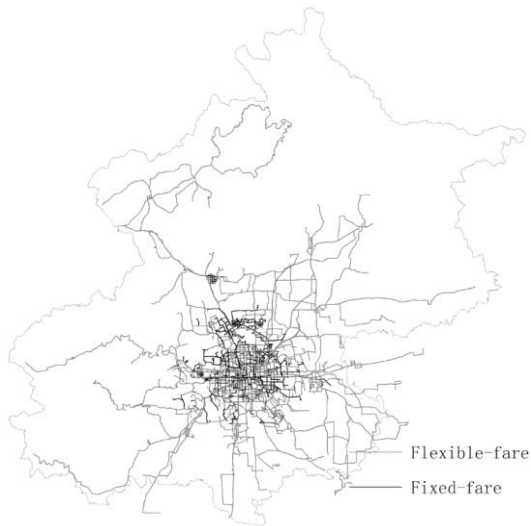
(a)
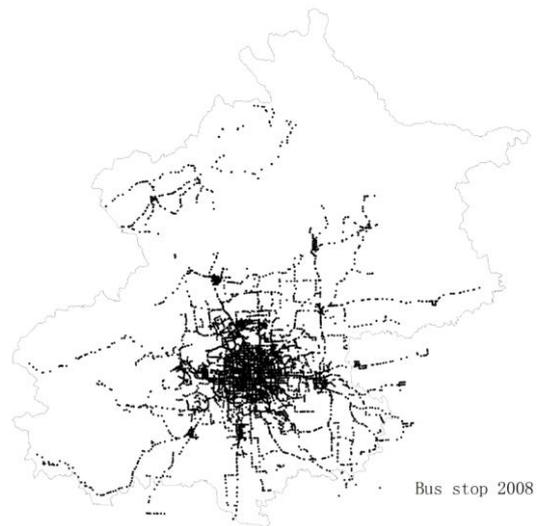
(b)
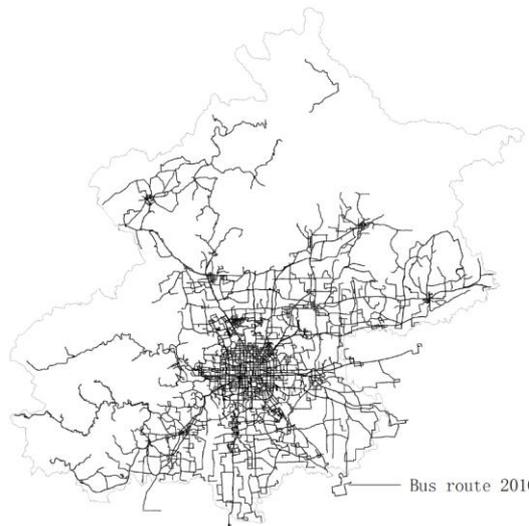
(c)
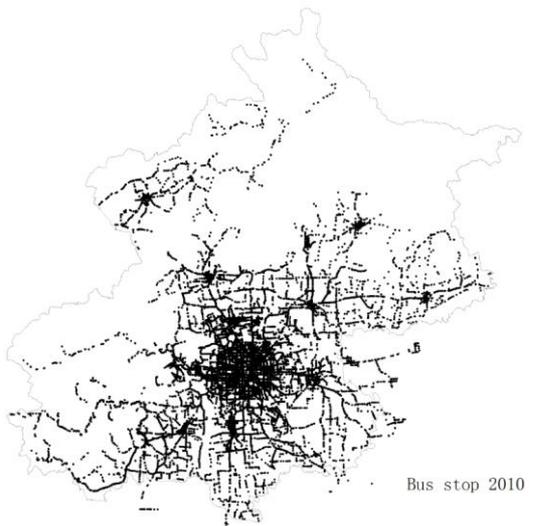
(d)
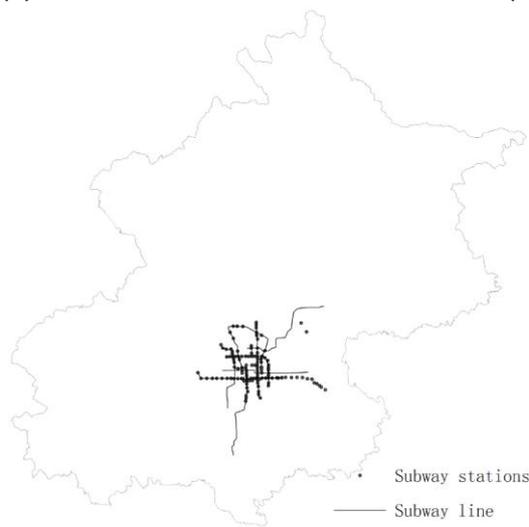
(e)
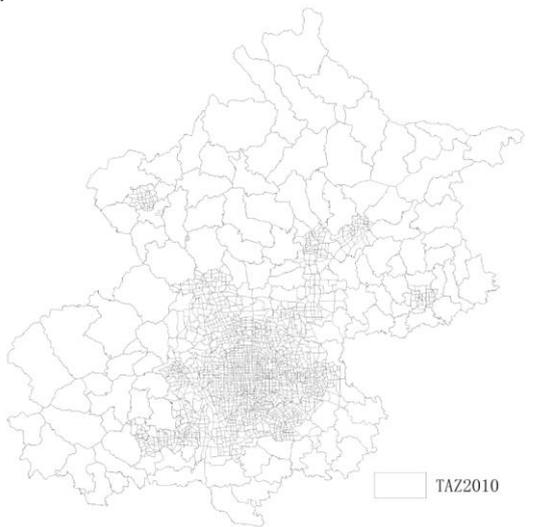
(f)

Figure 4 Bus routes in 2008 (a), bus stops in 2008 (b), bus routes in 2010 (c), bus stops in 2010 (d), subway lines and stations in 2010 (e), and traffic analysis zones in 2010 (TAZs) (f)

of the BMA. Note: All maps are from the Beijing Institute of City Planning. Some bus routes and stops are outside the BMA, as shown in a-d since some residents live outside the BMA and in adjacent towns in Hebei province.

### 3.2.3 Household travel survey in 2010

The 2010 Beijing Household Travel Survey data (the 2010 survey hereafter) was used for getting underprivileged resident samples from all surveyed residents in Beijing. This survey adopts a multistage sampling strategy with a target of 1% sampling rate. 1,085 out of 1,911 TAZs in the whole BMA are selected. In each TAZ, 10 to 50 households are selected to take a face-to-face interview. The final sample size is 46,900 households (116,142 residents) in the BMA. The 2010 Survey provides one-day travel diaries of all respondents, which gives commuting time for each employee (distance unavailable except the TAZs of Origin-Destination). The commuting mode is identified as the mode of the trip segment with the longest duration. We choose the mode of the highest mobility when there are two or more segments that have the same, longest duration. The 2010 Survey presents also household information including household structure, income, and residential location at the TAZ level, as well as personal information including gender, age, occupation, industry of employment, etc.

**We identify underprivileged residents with only metro/bus trips from the 2010 survey to reveal their spatial mobility.** These residents are limited to bus/metro users to link with SCD. We selected **4432** underprivileged residents (around 4%) by three steps. Note that since the 2010 survey income information is not convinced and we therefore only select underprivileged residents with extreme conditions (thus 4432 selected residents are with extreme underprivileged in terms of the selection standards we used). **First**, we select any residents in households with no car, annual income lower than 50,000 CNY, housing area smaller than 60 $m^2$, and with no Moto or e-bicycle. **Second**, from those residents selected by the first step, we select residents whose age is between 18 and 80 (including 18 and 80), and who are not students or foreigners, and have smartcard. **Third**, we limited the residents selected by the second step to those only having bus/metro trips and over one ride per day.

We then analyze their profile to gain knowledge of the underprivileged residents of Beijing via the 2010 survey. Out of all 3479 households with identified underprivileged residents, 699 households do not have at least an employed person, and average housing area of the 699 households is 44.1 $m^2$ (6-60). Regarding their housing condition, 68 live in affordable housing (should be 16% according to official report, Liao 2013), 935 as tenants, and 94 in relatives' or friends' housing, which shows that **affordable housing was still not common for underprivileged residents in 2010**. Among all identified underprivileged residents (1913 male and 988 with driving license), 3102 (70.0%) residents have jobs, in which 2958 are full time, and others part-time. For others, there are 929 retired, 418 housewives, and 362 unemployed. For their education, 7 hold postgraduate degrees, 121 hold bachelor degrees, and 2223 occupational education. They are 40 years old on average, 3348 (75.5%) have local *Hukou*. There are 128 staying in the current house for less than 6 months. Note that the 2010 survey only documents one-day travel diary and their socioeconomic attributes, but lacks housing and job change information, which is only available in long-term records.

Their one-day travel diaries are then extracted and explorative analyzed in detail. The trajectories of identified underprivileged residents are regarded as representative samples for depicting spatial mobility of underprivileged residents using metro/bus. **There are totally 9367 trips** (1757 walking trips not included) with 1337 metro trips (14.3%). We found that 4123 (93.0%) residents have two trips in a day, while others have more than two trips. We mapped all residents' housing and job places in space and found no significant characteristics of their jobs-housing places (see Figure 5), which means that it is not easy to define underprivileged residents simply with their housing and job places. In addition, there are 2827 (30.1%) commuting trips and 4405 (47.0%) back-home trips among all trips, denoting that they have limited recreational and social activities in their daily life.

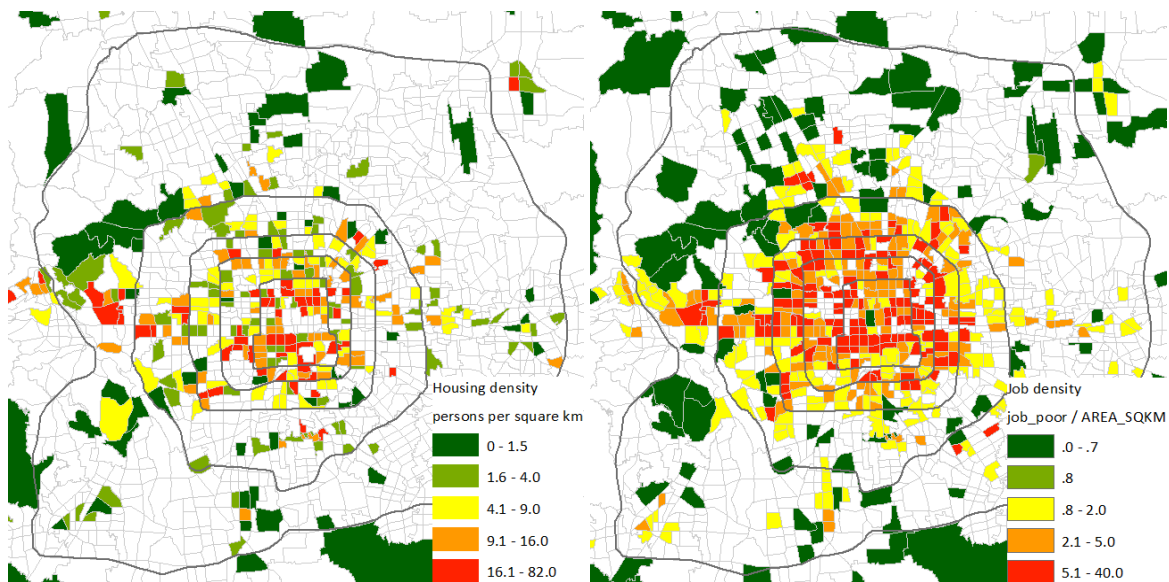

Figure 5 Housing and job density in the TAZ level

### *3.3 Underprivileged residents in Beijing and their mobility*

Better understanding the conditions of underprivileged residents in Beijing is necessary for answering our research questions with SCD. Generally, there are negligible underprivileged residents in Beijing. In 2010, the average household income for the bottom 20% households in terms of income was 16,988 CNY in Beijing (the Chinese poverty line is 2300 CNY in 2011), while the average household income for all households is 33,360 CNY (Beijing Municipal Statistics Bureau and NBS Survey Office in Beijing, 2011). Among them, there are 137,063 residents (23 million for the whole country) receiving pension from the local government (http://files2.mca.gov.cn/cws/201107/20110711152146727.htm), whose average monthly income was 366 CNY in Dec. 2010. A negligible number of them stay in urban villages in Beijing (Zheng et al, 2009). They could be classified into four social groups, the traditionally poor, fresh graduates (*Yizu*), unemployed adults, and rural migrants. The traditionally poor residents live in the old city and have low income due to the lack of work skill. Fresh graduates generally are from middle to low class universities, most of which are outside Beijing. The fresh graduates work in small private firms and cannot afford to rent a flat in downtown. Unemployed adults lose their jobs thus have no income or only have limited income from part-time jobs. Rural migrants, peasants from other provinces in China, come to Beijing for low-skilled jobs, like construction and household/business services. It should be noted that a small portion of the traditionally poor and unemployed adults can receive pension from the local government. The very limited pension, however, can only afford very basic living requirements in the capital city Beijing.

We developed several methods to gain knowledge on each type of underprivileged residents in Beijing, on who they are, how they are, and how their mobility is. First, we used literature review, planner interview, common sense of the first author of this manuscript as a local resident, as well as online small-scale survey via Sina Weibo, the Chinese Twitter widely used in mainland China. We analyzed each social group in terms of their conditions of housing, job, transportation and leisure, as highlighted by The Athens Charter.

**(1) Housing.** With the booming supply of constructed affordable housing projects by the government mainly since 2008 (construction starts during 2006-2007), more and more of them would assemble in these housing projects, 85% of which are located between the fourth and the sixth ring roads of Beijing. The application requirement for affordable housing is the size of current housing (less than 15 $m^2$ per person), annual household income (below 88,000 CNY), and *Hukou* conditions (limited to local urban residents). For the traditionally poor living in the center city, they have to live further away from central locations after the demolition of urban villages or with the gentrification process. For those who cannot afford buying a house, to keep close to their jobs and to save transportation expense, they tend to group-rent in downtown that several tenants squeeze into a room for low rent per person (a latest act in Beijing forbids this behavior). Some tenants living in small properties in suburb suffer from long distance commute by metro or bus. In addition, due to the informal rental market in Beijing, the tenants frequently change their housing places.

**(2) Job.** Most of them work for private enterprises rather than state-owned enterprises and do not have stable jobs, which is a common sense in Beijing (Gu et al, 2013). Since their jobs are not stable, a considerable proportion of them frequently change their jobs.

**(3) Transportation**. Due to the vehicle plate restriction, the increased taxi fare and parking fee, most of them have no car and travel by bus/metro/(e-)bicycle. Sheer cost of cars and petro is a more important factor to restrict their mobility. Most of them commute a long-distance to work, as revealed by Long and Thill (2013). For some extreme underprivileged residents like rural migrants, they still prefer buses and bicycles/electric bicycles than metro (although the metro fare in Beijing is very cheap as 2 CNY). They would only choose metro if they need to go somewhere so far that metro is the only sensible method. But for some fresh graduates, metro is very attractive.

**(4) Leisure.** Most of the underprivileged residents' activities are around their housing places. Fresh graduates tend to travel around metro stations with good accessibility, and they are the heavy users of online shopping, which has competitive price comparing with high-end shopping centers.

The above analysis has been summarized in Table 3, revealing that the places of residence and job of the underprivileged residents are generally not stable and change a lot over time (e.g. two years). Housing change, job change, and commute are essential for profiling and differentiating underprivileged residents in Beijing. We expect to find these aspects by using SCD.

Table 3: The profile of underprivileged residents in Beijing

| Type | | Housing | Job | Transport | Leisure |
|---|---|---|---|---|---|
| Traditional poor | Spatial | Downtown or in suburb after demolition | Local or far | | Local |
| | Non-spatial | Affordable housing | Small business/ unstable | (Electric) bicycle/bus | |
| Fresh graduates | Spatial | Downtown group rent or suburb rent | Business zones/downtown | Long commuting | Around subway stations |
| | Non-spatial | Affordable housing | Private firms/overload work/frequently change job | Metro/bus | Online shopping/buy sale |
| Unemployed adults | Spatial | | | | Local |
| | Non-spatial | Affordable housing | No job | (Electric) bicycle/bus | |
| Rural migrants | Spatial | Close to job | | | Local |
| | Non-spatial | Small property housing | Service/industry | (Electric) bicycle/bus | |

### *3.4 Most of frequent bus/metro riders in Beijing are underprivileged residents*

Almost all public transit smartcards are associated with only one cardholder during 2008-2010, which is our elementary assumption for the following analysis on SCD 2008 and 2010 [2]. That is, the user of a smartcard does not change during 2008-2010 in most cases, which especially applies to those frequent bus/metro riders. This is confirmed by Beijing Transportation Commission, BMAC, and our small-scale interview in Beijing. There are a considerable amount of public transit passengers in Beijing having several smartcards, some of which might be used by their relatives who visit Beijing for business or leisure. Further, it is not common for several passengers to share one smartcard in Beijing. In addition, some cardholders get their smartcards refunds when they do not need the smartcards or the smartcards do not work. Totally 20 CNY would be refund for each smartcard. As of 8 Jan 2014, there are 44 refund locations in Beijing, the few amounts of which cause a long queue at each location (generally over one hour according to the interviews). This problem leads to some passengers throwing the broken smartcards rather than refund in Beijing. The IDs of refunded smartcards would not be used anymore according to our inquiries with Beijing Planning Commission and BMAC.

---
[2] The lottery policy for getting a permit to buy a vehicle by individuals was issued on 1 Jan 2011. Therefore these middle-class residents who have no permit to buy a private car are not included in our 2008 and 2010 SCD.

With the proven assumption of one-card-for-one-passenger, we further argue that most of frequent bus/metro riders in Beijing were and are representatives of underprivileged residents in Beijing from the following aspects, although this has been discussed in the introduction section (the third paragraph). This argument needs further explanation since it is the elementary assumption of our paper to regard (most of) frequent bus/metro riders as underprivileged residents in Beijing and then to profile their typology and spatial mobility. This is supported by (1) a local household travel survey in 2010 (N=116,142); (2) a small-scale survey of transit riders in 2012 (N=709); (3) our interview with local residents in Beijing (N=46), and (4) the first author's local knowledge of Beijing.

**Proof 1, by the 2010 survey**

We have described the spatial mobility of extremely underprivileged residents in Beijing from the 2010 survey. That analysis is for revealing the mobility profile of them. To avoid potential bias in the revealed profile, we used very strict condition in selecting urban underprivileged residents in Beijing. In addition to the analysis, we released several standards to select frequent bus/metro riders, who are defined as not students, having a smartcard, and having over two (>=3) riding bus/metro ridings per day. Among all 1946 persons from 618 households identified as frequent riders, 1601 have Beijing *Hukou*, 1386 have no driving license, 913 have full time job, 47 have part-time job, 798 are retired, and others have no job. For the income of the 618 households, 417 are in the level 1, 166 are in the level 2, 191 are in the level 3, and 8 are in level 4. Moreover, 81 households have a car. According the aforementioned criteria of frequent bus/metro riders, they are economically underprivileged in terms of income, car ownership, as well as driving license ownership.

**Proof 2, by a small-scale survey**

In addition to the 2010 survey, we also conducted a small-scale survey on socio-economic attributes of cardholders in Shangdi and Qinghe sub-districts, Haidian district during September to December 2012. The survey was conducted by means of location sensing devices, an interactive survey website, questionnaires in person, and telephone. Totally 709 valid samples were retrieved (133 with valid smartcard ID) and each was associated with a valid card ID, socioeconomic attributes and corresponding travel diaries (one week). Among all 709 samples, there are 125 persons with at least one bus/metro trip per day. They are regarded as frequent bus/metro riders. Among 125 identified frequent riders, 80.8% residents' monthly income are lower than 6000 CNY, and most of them (50% of all) only in the 2001-4000 CNY level. Furthermore, among all 126 riders, 68 have no Beijing *Hukou* (a precondition for a stable job in Beijing), only 20 have driving licenses, and 31 have a car. Therefore, frequent bus/metro riders are also economically underprivileged.

**Proof 3, from interview with local residents in Beijing**

For getting further knowledge of frequent bus/metro riders in Beijing, we also interviewed local residents in Beijing. Most of them were friends and colleagues of the first author. We asked their, their relatives', friends' and colleagues' ideas on or experience with frequent bus/metro riders. Totally 34 residents among all 40 agree with the argument.

**Proof 4, from common sense of the first author who stayed in Beijing as a planner for over ten years**

Finally, according to our common sense on bus/metro frequent riders, those who ride bus/metro frequently for a long time, especially those with several transfers or a long distance, tend to be economically underprivileged.

Therefore, according to the proofs, most of frequent bus/metro cardholders (FCs) are underprivileged residents in Beijing. Some of FCs with a short commuting distance might be not underprivileged. We would consider this in the following method section via classifying FCs.

## 4 Method

Overall, the process for identifying and profiling underprivileged residents from SCD includes three steps. **First**, we identify housing and job places of cardholders using SCD in 2008 and 2010, respectively. The variations of housing and job places of cardholders are analyzed in detail. **Second**, those cardholders with housing places in both 2008 and 2010 are regarded as FRs. **Third**, we profile all FRs identified into 20 groups in terms of their housing place variation (change, no change), job place variation (change, no change, find a job, lose a job, and jobless all the time) during 2008-2010 and housing place in 2010 (within the fourth ring road or not) for deriving policy implications for practitioners. **Last**, the underprivileged degree of each FR is then evaluated.

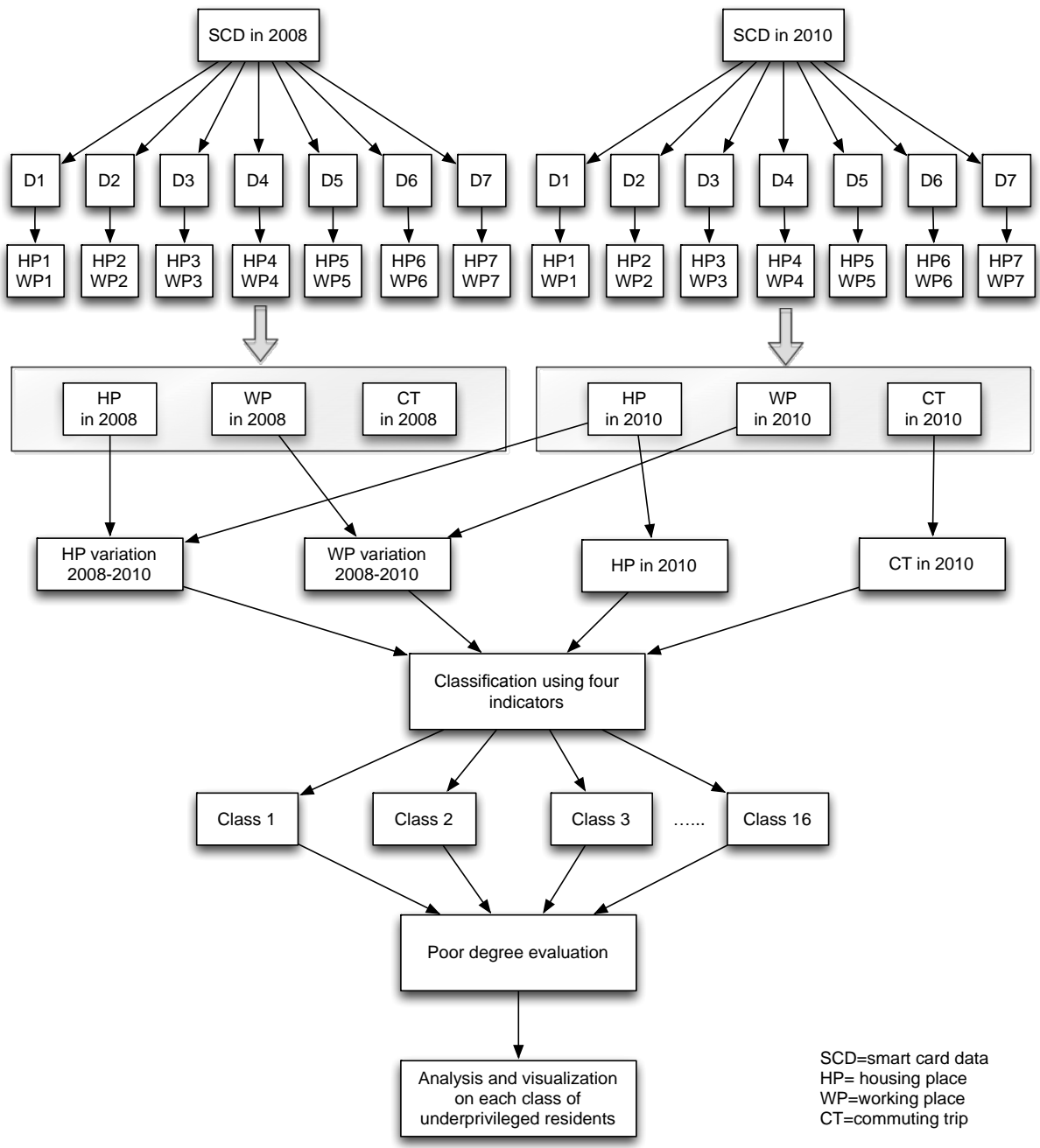

Figure 6 Flowchart for identifying and profiling underprivileged residents using SCD in 2008 and 2010

## *4.1 Housing and job place identification of all cardholders*

To identify a cardholder's workplace, we queried one-day data on a MS SQL Server and repeated the work for seven days based on these rules (see Long et al., 2013 for more details):
- The card type is not a student card;
- $D_j >= 6$ hours, where $D_j$ is the duration that a cardholder stays at place j, which is associated with all bus stops within 500 meters of one another;
- $j <> 1$, which means that j is not the first place in a weekday that the server records.

- The place where a cardholder visited most frequently in five weekdays will be defined as the final workplace of the cardholder in this study.

Similarly, we deduced from the data queries that a place would be a cardholder's housing place if the data meet these conditions:
- The cardholder has an identified workplace;
- The card type is not a student card;
- $D_h >= 6$ hours, where $D_h$ is the duration that a cardholder stays at place h, which is associated with all bus stops within 500 meters of one another;
- $F_h >= F_j$, where $F_h$ is the first and the most frequent place a cardholder starts a bus trip of a day within the week, $F_j$ is the trip frequency to or from j that the cardholder has.

We applied the methods for identifying housing and job places for SCD in 2008 and 2010, respectively. It should be mentioned that in the 2008 SCD identification, to ensure that we singled out commuters solely by bus, we only selected cardholders that had continuous bus swipes. That is, our study does not consider commuters who are multimodal and take bus and metro in identification of the 2008 SCD. This problem is absent from the 2010 SCD identification. Since the card ID keeps the same in 2008 and 2010, we could reveal a portion of cardholders' residential place change and job place change, which would be used in the following step. In this paper, we used the benchmark 2 km for variation in housing and working places. E.g., if a cardholder's housing place in 2008 is over 2 km desired distance to that in 2010, the cardholder's housing place would be assumed to change.

## *4.2 Underprivileged residents identification and classification*

Considering our discussion in the introduction section, Sections 3.3 and 3.4, we regard cardholders with identified housing places in both the 2008 SCD and the 2010 SCD as underprivileged residents due to the following reasons. First, bus/metro riders in Beijing are already less privileged. Second, a cardholder with identified housing place could guarantee that he/she used bus/metro frequently and had a stable housing place in the week. Third, continuing frequently riding bus/metro since 2008 to 2010 would be more underprivileged among all cardholders. We admit that small portions of identified underprivileged residents are not really poor. For these, we would figure it out via the following classification.

In accordance with the four urban functions by the Athens Charter, we classify all identified underprivileged residents into several groups in terms of three indicators, housing (changed or not), job (no change, changed, lose a job, find a job, jobless all the time), and housing location (within the fourth ring road or not). There would be 2*5*2=20 groups for all identified underprivileged residents as shown in Table 4. The size/ratio of each group could be calculated, and their trajectories in the 2010 SCD could be analyzed to profile their spatial mobility. Their social conditions could be profiled with the three indicators used for classification. Therefore, each group represents living styles of more underprivileged residents in Beijing.

Table 4: The profile of all types of underprivileged residents in Beijing (R4=The fourth ring road of Beijing)

| Group | Housing change | Job change | Housing location | # cardholders |
|---|---|---|---|---|
| 1 | Yes | Yes | Out of R4 | 6044 |
| 2 | Yes | Yes | Within R4 | 4055 |
| 3 | Yes | Find a job | Out of R4 | 13007 |
| 4 | Yes | Find a job | Within R4 | 8556 |
| 5 | Yes | Jobless | Out of R4 | 25129 |
| 6 | Yes | Jobless | Within R4 | 16760 |
| 7 | Yes | Lose the job | Out of R4 | 7635 |
| 8 | Yes | Lose the job | Within R4 | 4803 |
| 9 | Yes | No | Out of R4 | 672 |
| 10 | Yes | No | Within R4 | 421 |
| 11 | No | Yes | Out of R4 | 1728 |
| 12 | No | Yes | Within R4 | 886 |
| 13 | No | Find a job | Out of R4 | 3585 |
| 14 | No | Find a job | Within R4 | 1771 |
| 15 | No | Jobless | Out of R4 | 9023 |
| 16 | No | Jobless | Within R4 | 3918 |
| 17 | No | Lose the job | Out of R4 | 2374 |
| 18 | No | Lose the job | Within R4 | 1097 |
| 19 | No | No | Out of R4 | 761 |
| 20 | No | No | Within R4 | 349 |

# 5 Results

## *5.1 Identified FRs and their dynamics during 2008-2010*

The SCD identified results in 2008 and 2010 are listed in Table 5. There are totally 7.3 m cardholders having recorded trips in both 2008 and 2010 SCD. There are far more cardholders with housing and job places in 2010 than those in 2008, which attributes to the absent of metro records in the 2008 SCD.

Table 5 The profile of SCD identification results

| Indicator | | SCD2008 (*10$^3$) | SCD2010 (*10$^3$) | Both (*10$^3$) |
|---|---|---|---|---|
| Total cardholders | | 8.5 m | 10.9 m | 7.3 m |
| Cardholders with | Housing place | 1046 | 2121 | 113 |
| | Job place | 363 | 986 | 15 |
| | Commuting trip | 222 | 703 | 15 |
| | Commuting distance | 222 | 703 | 15 |
| | Commuting time | 222 | 420 | 9 |
| Average commuting distance (km) | | 8.2 | 10.6 | N/A |
| Average commuting time (min) | | 36.0 | 35.4 | |

Note: not all commuting trips in 2010 have the commuting time, since we only have the boarding time for distance-fare bus lines in the 2010 SCD.

There are 112,574 cardholders with identified housing places in both 2008 and 2010, which are defined as FRs and underprivileged residents in Beijing. Their housing place, workplace, and commuting dynamics during 2008-2010 are analyzed as shown in Table 6-8.

There are 77.4% FRs' housing places were changed during 2008-2010, and more FRs moved outward (most of them with over 5 km outward distance).

Table 6 Housing place dynamics of FRs during 2008-2010

| Housing place | | | # cardholders | Ratio (%) |
|---|---|---|---|---|
| Not changed | | | 25,492 | 22.6 |
| Changed | | | 87,082 | 77.4 |
| | | | 42,013 | 37.3 |
| | Inward (km) | 2-5 | 9,211 | 8.2 |
| | | 5-10 | 9,651 | 8.6 |
| | | 10-20 | 13,150 | 11.7 |
| | | >=20 | 10,001 | 8.9 |
| | | | 45,069 | 40.1 |
| | Outward (km) | 2-5 | 7,990 | 7.1 |
| | | 5-10 | 10,139 | 9.0 |
| | | 10-20 | 16,400 | 14.6 |
| | | >=20 | 10,540 | 9.4 |
| **Sum** | | | **112,574** | **100.0** |

The changes of FRs' workplaces are also significant during 2008-2010. Only 13.3% FRs kept working in 2008 and 2010. There are 14.1% FRs lost their jobs during 2008-2010. That is, each of them had an identified workspace in 2008 but not in 2010. We also found that 23.9% FRs got a job in 2010. Surprisingly there are 48.7% FRs kept jobless in 2008 and 2010. We assume most of them are retired residents, part-time workers and jobless residents (students not excluded from housing and job identification process).

Table 7 Workplace dynamics of FRs during 2008-2010

| Workplace | | | | # cardholders | Ratio (%) |
|---|---|---|---|---|---|
| Working | | | | 14916 | 13.3 |
| | Not changed | | | 2203 | 2.0 |
| | Changed | | | 12713 | 11.3 |
| | | Inward (km) | | 6142 | 5.5 |
| | | | 2-5 | 1444 | 1.3 |
| | | | 5-10 | 1893 | 1.7 |
| | | | 10-20 | 2071 | 1.8 |
| | | | >=20 | 734 | 0.7 |
| | | Outward (km) | | 6571 | 5.8 |
| | | | 2-5 | 1371 | 1.2 |
| | | | 5-10 | 2018 | 1.8 |
| | | | 10-20 | 2156 | 1.9 |
| | | | >=20 | 1026 | 0.9 |
| Losing job | | | | 15909 | 14.1 |
| Finding a job | | | | 26919 | 23.9 |
| Jobless | | | | 54830 | 48.7 |
| **Sum** | | | | **112,574** | **100.00** |

For those FRs with commuting trip identified in both 2008 and 2010, we analyzed the variation of their commuting distance. Totally 78.0% FRs' commuting distance increased from 2008 to 2010, indicating the commuting situation by bus/metro got worse during the period.

Table 8 Commuting distance variation of FRs (with commuting trips both in 2008 and 2010)

| Commuting distance in 2010 − that in 2008 (km) | # cardholders |
|---|---|
| >=20 | 436 |
| 10-20 | 1,885 |
| 5-10 | 2,266 |
| 2-5 | 2,419 |
| 0-2 | 2,647 |
| -2-0 | 1,984 |
| -5-(-2) | 1,416 |
| -10-(-5) | 1,069 |
| -20-(-10) | 622 |
| <=-20 | 172 |
| **Sum** | **14,916** |

We classified all these 112,574 FRs into 20 types according to the method discussed in Section 4. The total number of each type is shown in Table 4.

*5.2 Evaluation on underprivileged degree*

There are several separate approaches for evaluating the underprivileged degree for each group of underprivileged residents identified as follows. (1) Trajectory comparison. The trajectories of identified results of each type can be compared with those of urban poor only by bus/metro documented in the 2010 survey (4432 extremely poor), in terms of the similarity of trajectories. The more similar, the more underprivileged a type would be. (2) Residence context of housing places. This approach calculates the decent level of each platform (e.g. the spatial context of a platform like housing price or affordable housing projects or high-end amenities or high-income TAZs) in trajectories of identified cardholders by each model, and sum for each cardholder. The model with the least average decent level by cardholders would be selected as the best model for identifying urban poor using SCD. (3) Double check using the 2014 SCD. Average trip count in 2014 of all cardholders appearing in both the identified urban poor and the 2014 SCD (also one week) could be calculated. The model with a greater trip count total would be selected as the best model used in this paper. In this paper, we adopt the last approach for evaluating the underprivileged residents we identified.

Among all 112,574 FRs identified, 29,189 (25.9%) have at least one trip in a week in 2014. Their average trip count is 13.0 in 2014 and the standard deviation is 9.7. Those 29,189 residents are regarded as the most underprivileged FRs, and their underprivileged degree increases with the trip count in 2014. For those majorities not appearing in the 2014 SCD, there are several reasons, (1) their economic condition has improved and become private car drivers; (2) they do not use the previous cards but still ride bus by other cards. This condition is not dominating in Beijing according to our local knowledge; (3) they moved out of Beijing, which is common with increasing living cost in the city. In sum, the dynamics of identified FRs was huge during 2010-2014, which is normal when comparing with those middle-class residents in Beijing.

# 6 Conclusions and discussion

This article explores on using public transit smartcard data during 2008-2010 for identifying and profiling economically underprivileged residents in Beijing. The ground truth of underprivileged residents in Beijing has been elaborated from their housing, job, transport and leisure aspects by various means. The mobility of them has been disclosed by analyzing economically underprivileged residents documented in the 2010 Beijing household travel survey. Considering Chinese situation, we then regarded these frequent public transit passengers as underprivileged residents, which have been proved by several avenues. This empirical study includes the following aspects. **First**, we extracted rules from the 2010 survey for identifying housing and job places for cardholders in the 2008 and 2010 SCD, respectively. The inherent smartcard IDs during 2008-1010 enabled us to detect shared cardholders' housing, job and commuting dynamics in the period. **Second**, we regarded these cardholders with identified housing place in both 2008 and 2010 as underprivileged residents in Beijing. We then profile them into 20 groups considering their housing place variation (change, no change), job place variation (change, no change, find a job, lose a job, and jobless all the time) during 2008-2010 and housing location in 2010 (within the fourth ring road or not). The mobility pattern of each group has been profiled in detail. **Last**, the underprivileged degree of cardholders in each group were estimated and compared with each other.

The methodological contributions of this paper are as follows. **First**, we use two-year-spanning immediate big data for understanding long-term urban dynamics, which is not possible by short-term data and only available when big data are accumulated (Batty, 2013). **Second**, we propose a promising solution on extracting rules from conventional travel surveys and using them for identifying interested information from big data. The simple and straightforward method has its potential in applying to other cities with access to SCD. **Third**, a large number of underprivileged residents in Beijing are for the first time profiled in terms of their housing, job and spatial mobility, which is not easy by conventional surveys in the background that Chinese people dislike disclosing their social condition and detailed spatial mobility (the gap of average household income in the 2010 survey and actual is an evidence). In sum, the urban poor analysis is not easy by other types of big data, and SCD have advantage for this issue in China considering the special situation.

In addition to our methodological contributions, we see opportunities on its application in practices. **First**, the revealed profile of underprivileged residents in terms of their housing, job and mobility can be regarded as references to the location choice of poor-aware facilities/amenities, like affordable housing and related job markets. They can also be checked for introducing jobs, including them in ticket fare audit meetings as the party of low income. **Second**, the identification results can be referred in the process of social welfare application like affordable houses and monetary pensions. That is, whether an applicant qualifies the social welfare standard can be double-checked by his/her trips using SCD, in addition to stated or reported indicators. **Third**, the 60% public transit fare discount on using smartcard has been applied to all passengers since the use of smartcards in 2005. There are appeals on raising the fare due to unaffordable operation cost by the local government. In this situation, one possible solution is that public transit fare subsidization can go to these identified underprivileged residents, who are in great need of fare discount, comparing to others. **Fourth**, the profiled mobility patterns of identified underprivileged residents can be addressed in revising public transit planning and design.

While admitting the merits of our study, there are several potential biases to be addressed in our future research. **First**, metro records are not included in the 2008 SCD, resulting in those cardholders commuting by metro from home are absent from identified underprivileged residents. **Second**, how the underprivileged residents identified from SCD are representative of all is an unavoidable question, as discussed by Liu et al (2013) for social network users and Sun et al (2012) for train passengers. For instance, underprivileged residents with no bus/metro trips are not accounted in this study. These residents by e-bicycles are more underprivileged by our common sense in Beijing, which could be complemented by the 2010 survey. In addition, we are extending our data analysis from 2008-2010 to 2008-2013 for understanding a longer term urban dynamics like residential and job location choice.

**Acknowledgments:** The first author would like to acknowledge the financial support of the National Natural Science Foundation of China (No. 51408039), China Scholarship Council (No.201209110066), and the 12th Five-year National Science Supported Planning Project of China (2012BAJ05B04). Any errors and inadequacies of the paper remain solely the responsibility of the authors.## References

Amini, A., Kung, K., Kang, C., Sobolevsky, S., & Ratti, C. (2014). The Impact of Social Segregation on Human Mobility in Developing and Urbanized Regions.arXiv preprint arXiv:1401.5743.

Batty, M. (2012). Editorial. *Environment and Planning B: Planning and Design*, *39*, 191–193.

Batty, M., (2013) Urban Modelling: A Progress Report. In Symposium on Applied Urban Modelling (AUM 2013), University of Cambridge.

Beijing Municipal Statistics Bureau, NBS Survey Office in Beijing (2011). Beijing statistical yearbook 2011. Beijing: China Statistics Press.

Beijing Transportation Research Center, 2011. *Beijing Transportation Annual Report 2009*. Unpublished Official Report. (In Chinese). Retrieved from http://www.bjtrc.org.cn/InfoCenter/NewsAttach//aeb7c878-d31e-4f08-982f-3c17c717c87b.pdf. Accessed on 2013-9-21.

Blythe, P. (2004). Improving public transport ticketing through smart cards. *Proceedings of the Institute of Civil Engineers, Municipal Engineers*, *157*, 47–54.

Chen, G., Gu, C., & Wu, F. (2006). Urban poverty in the transitional economy: a case of Nanjing, China. Habitat International, 30(1), 1-26.

Cheng, Z., Caverlee, J., Lee, K., & Sui, D. Z. (2011). Exploring Millions of Footprints in Location Sharing Services. ICWSM, 2011, 81-88.

Cho, E., Myers, S. A., & Leskovec, J. (2011). Friendship and mobility: user movement in location-based social networks. In Proceedings of the 17th ACM SIGKDD international conference on Knowledge discovery and data mining (pp. 1082-1090). ACM.

Fan, C. C., Sun, M., & Zheng, S. (2011). Migration and split households: a comparison of sole, couple, and family migrants in Beijing, China. Environment and Planning-Part A, 43(9), 2164.

Frias-Martinez, V., Virseda-Jerez, J., & Frias-Martinez, E. (2012). On the relation between socio-economic status and physical mobility. Information Technology for Development, 18(2), 91-106.

Gu, Y., Xu, Y., Zheng, S., (2013) Job Uncertainty, Residential Location Choices and Commuting Patterns in Beijing. Working paper.

Gong, Y., Liu. Y., Lin, Y., Yang, J., Duan, Z., & Li, G. (2012). Exploring spatiotemporal characteristics of intra-urban trips using metro smartcard records. *Proceedings of Geoinformatics*. Hong Kong.

Goodchild, M. F. (2007). Citizens as sensors: The world of volunteered geography. *GeoJournal*, *69*(4), 211–221.

Isaacman, S., Becker, R., Cáceres, R., Kobourov, S., Martonosi, M., Rowland, J., & Varshavsky, A. (2011). Identifying important places in people's lives from cellular network data. In Pervasive Computing (pp. 133-151). Springer Berlin Heidelberg.

Jang, W. (2010). Travel time and transfer analysis using transit smart card data. *Transportation Research Record: Journal of the Transportation Research Board*, *2144*, 142–149.23